\documentclass[aip,amsmath,amssymb,lengthcheck,showpacs]{revtex4-1}
\usepackage{graphicx}
\def\vec#1{\mbox{\boldmath $#1$}}
\newcommand{\vd}{\mathcal{D}}

\begin{document}

\title{Dynamical Heterogeneity in a Highly Supercooled Liquid under a Sheared Situation}

\author{Hideyuki Mizuno}
\email{h-mizuno@cheme.kyoto-u.ac.jp}
\author{Ryoichi Yamamoto}
\email{ryoichi@cheme.kyoto-u.ac.jp}
\affiliation{Department of Chemical Engineering, Kyoto University, Kyoto 615-8510, Japan}
\affiliation{CREST, Japan Science and Technology Agency, Kawaguchi 332-0012, Japan}

\date{\today}

\begin{abstract}
In the present study, we performed molecular-dynamics simulations and investigated dynamical heterogeneity in a supercooled liquid under a steady shear flow.
Dynamical heterogeneity can be characterized by three quantities: the correlation length $\xi_4(t)$, the intensity $\chi_4(t)$, and the lifetime $\tau_\text{hetero}(t)$.
We quantified all three quantities by means of the correlation functions of the particle dynamics, i.e., the four-point correlation functions, which are extended to the sheared condition.
Here, to define the local dynamics, we used two time intervals $t=\tau_\alpha$ and $\tau_{\text{ngp}}$; $\tau_\alpha$ is the $\alpha$-relaxation time, and $\tau_{\text{ngp}}$ is the time at which the non-Gaussian parameter of the Van Hove self-correlation function is maximized.
We discovered that all three quantities ($\xi_4(t)$, $\chi_4(t)$, and $\tau_\text{hetero}(t)$) decrease as the shear rate $\dot{\gamma}$ of the steady shear flow increases.
For the time interval $t=\tau_\alpha$, the scalings $\xi_4(\tau_\alpha) \sim \dot{\gamma}^{-0.08}$, $\chi_4(\tau_\alpha) \sim \dot{\gamma}^{-0.26}$, and $\tau_\text{hetero}(\tau_\alpha) \sim \dot{\gamma}^{-0.88}$ were obtained.
The steady shear flow suppresses the heterogeneous structure as well as the lifetime of the dynamical heterogeneity.
In addition, our results demonstrated that the $\alpha$-relaxation time $\tau_\alpha$ dependences of three quantities coincide with those at equilibrium.
This means that all three quantities of the dynamical heterogeneity can be mapped onto those in the equilibrium state through $\tau_\alpha$.
\end{abstract}


\maketitle

\section{Introduction}
When liquids are cooled toward the glass transition temperature $T_g$, the dynamics not only drastically slow down, but also become progressively more heterogeneous \cite{muranaka_1994,hurley_1995,perera_1999,widmer_2004,doliwa_2002,donati_1998,Marcus1999,kegel_2000,weeks_2000}.
``Dynamical heterogeneity" has attracted much attention because it may lead to a better understanding of slow dynamics \cite{ediger_2000,dhbook}.
Dynamical heterogeneity is characterized by three quantities: the correlation length $\xi_4$, the intensity $\chi_4$, and the lifetime $\tau_{\text{hetero}}$ \cite{glotzer_2000,doliwa2000,lacevic_2003,berthier_2004,toninelli_2005,chandler_2006,stein_2008,karmakar_2010,flenner_2010,flenner_2011,flenner_2004,leonard_2005,kim_2009,kim_2010,mizuno_2010,mizuno2_2010}.
The correlation length $\xi_4$ and the intensity $\chi_4$ characterize the static properties of the dynamical heterogeneity and can be quantified by means of the static structure factor of the particle dynamics, i.e., the so-called four-point correlation function \cite{glotzer_2000,doliwa2000,lacevic_2003,berthier_2004,toninelli_2005,chandler_2006,stein_2008,karmakar_2010,flenner_2010,flenner_2011}.
The lifetime $\tau_{\text{hetero}}$ measures the dynamical properties of the dynamical heterogeneity, and we can evaluate the lifetime by the time decay of the correlation function of the particle dynamics \cite{flenner_2004,leonard_2005,kim_2009,kim_2010}.
In our previous studies \cite{mizuno_2010,mizuno2_2010}, we have consistently quantified these three quantities (correlation length, intensity, and lifetime) from a single order parameter and its correlation functions.
The length scale and the time scale of the dynamical heterogeneity tend to diverge as the temperature approaches the glass transition point.

Dynamical heterogeneity has also been investigated for a sheared situation \cite{yamamoto1_1998,tsamados_2010,nordstrom_2011,heussinger_2010,heussinger2_2010}.
When applying a steady shear flow to supercooled liquids, marked shear-thinning can be observed \cite{varnik_2006,miyazaki_2004,bessel_2007}.
In the sheared situation, the dynamics become not only much faster but also more homogeneous; therefore, the shear flow suppresses the dynamical heterogeneity.
Refs. \cite{tsamados_2010,heussinger_2010,nordstrom_2011} found that the intensity $\chi_4$ of the dynamical heterogeneity decreases with an increasing shear rate $\dot{\gamma}$ as $\chi_4 \sim \dot{\gamma}^{-\mu}$ with $\mu=0.3$ \cite{tsamados_2010,nordstrom_2011} or $\mu=0.4-0.6$ \cite{heussinger_2010}.
To the best of our knowledge, the lifetime $\tau_{\text{hetero}}$ of the dynamical heterogeneity has not yet been examined for the sheared situation.
Dynamical heterogeneity may play an important role in the drastic change of the dynamics due to not only the decreasing temperature but also the increasing shear rate.
However, the role of dynamical heterogeneity remains unclear and is an important question.

Even in a strongly sheared state, supercooled liquids exhibit an almost isotropic structure and dynamics that can be captured via two-point correlation functions \cite{yamamoto1_1998,miyazaki_2004,bessel_2007}.
This is in marked contrast to common complex fluids, such as polymer solutions and colloidal suspensions, which exhibit anisotropic dynamics or structural changes that are induced by the shear flow \cite{rheology,onuki_1997,phasetransition}.
The isotropic feature of supercooled liquids strongly supports the possibility of the ``mapping concept", that is, that the sheared non-equilibrium state can be mapped onto the quiescent equilibrium state.
In fact, previous studies \cite{yamamoto1_1998,yamamoto_1998} have demonstrated that the shear viscosity $\eta$ and the diffusion constant $D$ in the sheared state can be mapped onto those in the equilibrium state through the $\alpha$-relaxation time $\tau_\alpha$.
The research by Ref. \cite{haxton_2007} also reported that the shear stress and the inherent structure energy can be mapped through the effective temperature $T_{\text{eff}}$, which is measured from the relation between the static linear response and the variance of the pressure.
More interestingly, there are some results regarding the mapping of the dynamical heterogeneity.
An earlier study \cite{yamamoto1_1998} examined the heterogeneous structure of the bond breakages under the sheared situation and found that the correlation length of the bond breakages is described as a function of only the $\alpha$-relaxation time.
Furthermore, recent research \cite{nordstrom_2011} has found that the intensity $\chi_4$ can be scaled as $n^{\ast} \sim (\dot{\gamma} \Delta \phi^{4})^{-0.3}$, where $\dot{\gamma}$ is the shear rate of the shear flow, and $\Delta \phi$ is the distance from the random close-packing volume fraction.
This result means that the intensity can be mapped through a single parameter $\dot{\gamma} \Delta \phi^{4}$.

In the present study, we examined the dynamical heterogeneity in the sheared non-equilibrium situation using molecular-dynamics (MD) simulations.
Three quantities (correlation length, intensity, and lifetime) characterizing the dynamical heterogeneity were quantified by means of the four-point correlation functions.
We found the behaviors of decreasing of not only the correlation length $\xi_4$ and the intensity $\chi_4$ but also the lifetime $\tau_{\text{hetero}}$ with increasing the shear rate $\dot{\gamma}$.
Furthermore, we also examined the validity of the mapping concept for all three quantities of the dynamical heterogeneity.

The mapping concept can lead to the concept of an effective temperature, which is also interesting and important for glassy systems.
For glassy systems, it has been suggested that when considering the long time scale, the equilibrium form of the FDT holds well with the temperature $T$, which is replaced by a different value denoted as $T_{\text{eff}}$, i.e., the effective temperature \cite{berthier_2000,berthier_2002,ono_2002,makse_2002,hern_2004,potiguar_2006,Martens_2009,kruger_2010,zhang_2011}.
If the concept of an effective temperature is valid, then the effective temperature plays a role as the temperature in the non-equilibrium situation, and the sheared non-equilibrium state can correspond to the equilibrium state with the effective temperature $T_{\text{eff}}$.
However, there are difficulties regarding the mapping concept and the effective temperature.
A recent numerical simulation \cite{furukawa_2009} found out the anisotropic dynamics in a sheared supercooled liquid by the four-point correlation function.
This finding indicates that the sheared state cannot completely be mapped onto the equilibrium state.
Furthermore, the validity of the value $T_\text{eff}$, which is calculated from the relation between the response function and its associated correlation function, remains uncertain.
Although several different observables yield a common value of $T_\text{eff}$ \cite{berthier_2002,ono_2002,makse_2002,potiguar_2006}, some observables do not provide consistent values of $T_\text{eff}$ \cite{hern_2004,zhang_2011}.
Therefore, only one value of $T_{\text{eff}}$ may not completely describe the non-equilibrium state.
Despite these difficulties, the mapping concept and the effective temperature are very attractive and interesting.

The paper is organized as follows.
In Sec. \ref{3simmodel}, we briefly review our MD simulation and the correlation functions of the particle dynamics.
In this section, we extend the correlation functions to the present sheared condition.
In Sec. \ref{3result}, we present our results.
First, we show the shear rate dependences of the correlation length, the intensity, and the lifetime, and then we demonstrate whether the mapping concept is valid for these three quantities .
In Sec. \ref{3conclusion}, we summarize our results.

\section{Simulation model and correlation functions of particle dynamics} \label{3simmodel}
\subsection{Simulation model}
We performed MD simulations in three dimensions.
Our simulation model is a mixture of two different size atomic species, 1 and 2.
The particles interact via the soft-sphere potential $v_{a b}(r)= \epsilon (\sigma_{a b}/r)^{12}$ with $\sigma_{a b} = (\sigma_a + \sigma_b)/2$, where $r$ is the distance between two particles, $\sigma_a$ is the particle size, and $a,b \in 1,2$.
The interaction was truncated at $r=3 \sigma_{a b}$.
We took the mass ratio to be $m_2/m_1=2$ and the size ratio to be $\sigma_2/\sigma_1=1.2$.
This diameter ratio avoided system crystallization and ensured that an amorphous supercooled state occurred at low temperatures \cite{miyagawa_1991}.
As in our previous study \cite{mizuno_2010}, we used two systems: a small system with $N_1 = N_2 =5 \times 10^3$ ($N=N_1+N_2=10^4$) particles and a large system with $N_1 = N_2 =5 \times 10^4$ ($N=N_1+N_2=10^5$) particles.
We quantified the correlation length $\xi_{4}(t)$ and the intensity $\chi_{4}(t)$ with the large system, and we quantified the lifetime $\tau_{\text{hetero}}(t)$ with the small system.
In the following, space, time, and temperature were measured by $\sigma_1$, $\tau_0 = (m_1 \sigma_1^2/\epsilon)^{1/2}$, and $\epsilon/k_B$, respectively.
The particle density was fixed at a value of $\rho = (N_1 + N_2)/V = 0.8$, and the system length was $L=V^{1/3} = 23.2$ and $50.0$ for the small and large systems, respectively.
The temperature was set as $T=0.772,\ 0.473,\ 0.352,\ 0.306,\ 0.267$, and $0.253$.
Note that the freezing point of the corresponding one-component model is approximately $T=0.772$ \cite{miyagawa_1991}.
At $T=0.253$, the system is in a highly supercooled state.
We applied a steady shear flow on our system with the Lees-Edwards periodic boundary condition \cite{nonequ}.
We integrated the so-called SLLOD equations of motion with the Lees-Edwards periodic boundary condition, and the temperature was maintained at a desired value using a Gaussian constraint thermostat.
Here, we set the $x$ axis and the $y$ axis along the flow direction and the velocity gradient direction of the steady shear flow, respectively.
The mean velocity profile is $\langle \vec{v} \rangle = \dot{\gamma} y \vec{e}_x$, where $\vec{e}_x$ is the unit vector in the $x$ direction.
The details of this simulation model can be found in previous studies \cite{yamamoto1_1998,miyazaki_2004}.

\begin{figure}[t]
\begin{center}
\includegraphics[scale=1]{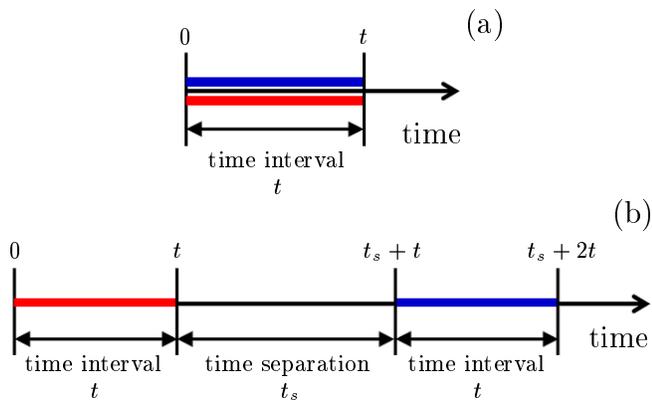}
\end{center}
\vspace*{-3mm}
\caption{
Schematic illustration of the time configuration of the correlation functions of the particle dynamics: 
(a) $S_4(q,t)$
(b) $S_\vd(q,t_s,t)$.
}
\label{ts}
\end{figure}

\subsection{Correlation functions of particle dynamics}
In our previous studies \cite{mizuno_2010, mizuno2_2010}, we investigated dynamical heterogeneity in the unsheared equilibrium state.
With the correlation functions of the particle dynamics, we quantified the correlation length $\xi_4(t)$, the intensity $\chi_4(t)$, and the lifetime $\tau_{\text{hetero}}(t)$, three of which characterize the static and dynamic properties of dynamical heterogeneity.
In those studies \cite{mizuno_2010, mizuno2_2010}, we defined an order parameter $\delta \hat{\vd}_1(\vec{r},t_0,t)$ and its Fourier component $\delta \vd_1(\vec{q},t_0,t)$ as the local fluctuations in the particle mobility.
In the present study, we extended $\delta \hat{\vd}_1(\vec{r},t_0,t)$ and $\delta \vd_1(\vec{q},t_0,t)$ to the sheared condition as follows \cite{furukawa_2009}.
A position $\vec{r}(t_0) = (x(t_0),y(t_0))$ on the reference frame at a time $t_0$ moves to $\vec{r}(t_0) + \dot{\gamma} t y(t_0) \vec{e}_x$ at a time $t_0 + t$ by the mean shear flow.
Therefore, we must define the mobility $a_{1j}^2(t_0,t)$ of the particle $j$ by
\begin{equation}
a_{1j}^2(t_0,t) = \frac{ [\vec{r}_{1j}(t_0+t)-\vec{r}_{1j}(t_0)-\dot{\gamma} t y_{1j}(t_0) \vec{e}_x]^2 }{ \langle [\vec{r}_{1j}(t_0+t)-\vec{r}_{1j}(t_0)-\dot{\gamma} t y_{1j}(t_0) \vec{e}_x]^2 \rangle }.
\end{equation}
Note that when $a_{1j}^2(t_0,t)\ge 1$ ($a_{1j}^2(t_0,t)<1$), the particle $j$ moves more (less) than the mean value of the single-particle displacement, i.e., the particle $j$ is mobile (immobile).
Using this definition of particle mobility, we extended $\delta \hat{\vd}_1(\vec{r},t_0,t)$ and $\delta \vd_1(\vec{q},t_0,t)$ to the sheared condition as
\begin{equation}
\begin{aligned}
& \delta \hat{\vd}_1(\vec{r},t_0,t) = \sum_{j=1}^{N_1} [a_{1j}^2(t_0,t)-1] \delta(\vec{r} - \vec{R}_{1j}(t_0,t)), \\
& \delta \vd_1(\vec{q},t_0,t) = \sum_{j=1}^{N_1} [a_{1j}^2(t_0,t)-1] \exp[-i\vec{q} \cdot \vec{R}_{1j}(t_0,t) ],
\end{aligned}
\end{equation}
where $\vec{R}_{1j}(t_0,t) = ( \vec{r}_{1j}(t_0) + \vec{r}_{1j}(t_0+t) )/2$.

Using the order parameter $\delta \vd_1(\vec{q},t_0,t)$, we define the spatial correlation function $S_4(q,t)$ of the particle dynamics as
\begin{equation}
S_{4}(q,t) = \frac{1}{N_1} \langle \delta \vd_1(\vec{q},0,t) \delta \vd_1(-\vec{q},0,t) \rangle, \label{sdqst}
\end{equation}
where the $\langle \cdots \rangle$ represents the ensemble average over time $0$ and the angular components of the wave vector $\vec{q}$.
The function $S_4(q,t)$ represents the spatial correlation of the particle dynamics in the time interval $[0,t]$.
The time configuration of $S_{4}(q,t)$ is schematically illustrated in Fig. \ref{ts}(a).
We are able to quantify the correlation length $\xi_{4}(t)$ and the intensity $\chi_{4}(t)$ by fitting $S_4(q,t)$ to the Ornstein-Zernike (OZ) form at small wavenumbers $q$ \cite{yamamoto1_1998,lacevic_2003},
\begin{equation}
S_{4}(q,t) = \frac{\chi_{4}(t)}{1 + q^2 \xi_{4}(t)^2}. \label{OZd}
\end{equation}
Note that to obtain accurate values of $\xi_{4}(t)$ and $\chi_{4}(t)$, $S_{4}(q,t)$ must be fitted to the OZ form in the range of $q \xi_{4}(t) < 1.5$ \cite{flenner_2010,flenner_2011}.
We used a large system with $10^5$ particles and carefully fitted $S_{4}(q,t)$ to the OZ form.

In addition, we also define the time correlation function $S_{\vd}(q,t_s,t)$ of the particle dynamics as
\begin{equation}
S_{\vd}(q,t_s,t) = \langle \delta \vd_1(\vec{q},t_s+t,t) \delta \vd_1(-\vec{q},0,t) \rangle, \label{sdqt}
\end{equation}
where the $\langle \cdots \rangle$ represents the ensemble average over the initial time $0$ and the angular components of the wave vector $\vec{q}$.
The function $S_{\vd}(q,t_s,t)$ represents the correlation of the particle dynamics between two time intervals $[0,t]$ and $[t_s+t,t_s+2t]$, where the value $t_s$ is the time separation.
The time configuration of $S_{\vd}(q,t_s,t)$ is schematically illustrated in Fig. \ref{ts}(b).
Here, when calculating the ensemble average over the wave vector $\vec{q}$, we took only $\vec{q}$ perpendicular to $x$ direction, i.e., $\vec{q}$ perpendicular to the shear flow direction, to eliminate the time decay due to the shear flow.
As the time separation $t_s$ increases, $S_{\vd}(q,t_s,t)$ decays in the stretched exponential form,
\begin{equation}
\begin{aligned}
\frac{S_{\vd}(q,t_s,t)}{S_{\vd}(q,0,t)}
\sim \exp\left( - \left( \frac{t_s}{\tau_{h}(q,t)} \right)^c \right),
\end{aligned} \label{relaxsd}
\end{equation}
where $\tau_{h}(q,t)$ is the wavenumber-dependent relaxation time of $S_{\vd}(q,t_s,t)$.
The lifetime $\tau_{\text{hetero}}(t)$ was determined as $\tau_{\text{hetero}}(t) = \tau_{h}(q,t)$ at $q=0.38$.
We used a small system with $10^4$ particles to obtain the lifetime $\tau_{\text{hetero}}(t)$.

As we mentioned previously, recent research \cite{furukawa_2009} has indicated that the correlation functions of the particle dynamics (the four-point correlation functions) depend on the direction of the wave vector $\vec{q}$ in the sheared situation, although the two point correlation functions hardly exhibit any anisotropy.
Therefore, three quantities ($\xi_4(t)$, $\chi_4(t)$, and $\tau_{\text{hetero}}(t)$) also depend on the direction of $\vec{q}$.
In the present study, we did not examine the directional dependence of the dynamical heterogeneity, even though it is interesting and important.
Rather, we examined the dynamical heterogeneity and its three quantities averaged over the direction of the wave vector $\vec{q}$.

\begin{figure}[t]
\begin{center}
\includegraphics[scale=1]{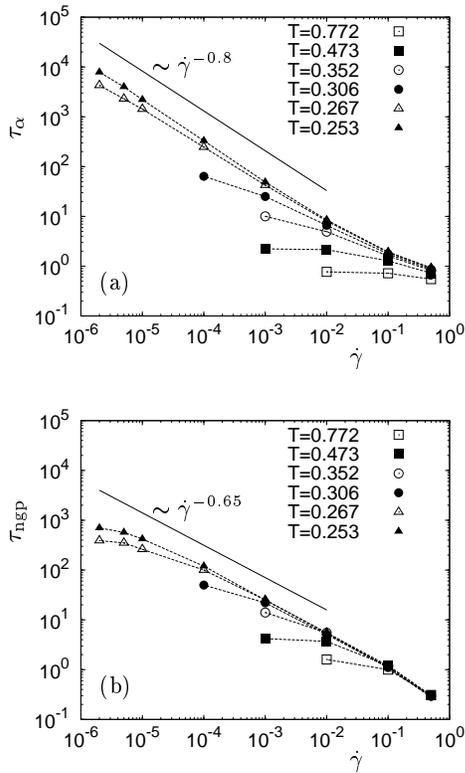}
\end{center}
\vspace*{-3mm}
\caption{The shear rate dependence of (a) $\tau_\alpha$ and (b) $\tau_{\text{ngp}}$ for various temperatures.
We used these two time intervals to define the particle dynamics.}
\label{taua}
\end{figure}

\section{Results}  \label{3result}
To define the particle dynamics, we used two time intervals as in our previous studies \cite{mizuno_2010,mizuno2_2010}.
One is the $\alpha$-relaxation time $\tau_\alpha$ defined by $F_s(k_m,\tau_\alpha)=e^{-1}$.
Here, $F_s(k,t)$ is the self-part of the density correlation function for Particle Species 1, which is defined as $F_s(k,t) = 1/N_1 \left< \sum_{j=1}^{N_1} e^{-i \vec{k} \cdot [ \vec{r}_{1j}(t) - \vec{r}_{1j}(0) - \dot{\gamma} t y_{1j}(t) \vec{e}_x ]} \right>$, and $k_m = 2\pi$ is the first peak wave number of the static structure factor.
We note that, as already shown in Refs. \cite{yamamoto1_1998,miyazaki_2004,bessel_2007}, any angular dependence in $F_s(\vec{k},t)$ is barely notable despite the presence of the shear flow in the $x$ direction.
Figure \ref{taua}(a) shows $\tau_\alpha$ as a function of the shear rate $\dot{\gamma}$.
The value $\tau_\alpha$ monotonically decreases with increasing $\dot{\gamma}$ as $\tau_\alpha \sim \dot{\gamma}^{-\nu}$ with $\nu \simeq 0.8$ \cite{yamamoto1_1998,miyazaki_2004,bessel_2007}.
Note that the marked shear thinning occurs in the region of the shear rate that we considered.

The other time interval is the time $\tau_\text{ngp}$ at which non-Gaussian parameter $\alpha_2(t)$ of the Van Hove self-correlation function is maximized.
Here, we introduce a displacement $\Delta\vec{r}'_{1j}(t)$ of the particle $j$, which is defined as $\Delta\vec{r}'_{1j}(t) = \vec{r}_{1j}(t)  - \vec{r}_{1j}(0) - \dot{\gamma} \int_0^t ds y_{1j}(s) \vec{e}_x$, in which the contribution from the convective transport by the average shear flow is subtracted \cite{yamamoto1_1998}. By using this displacement, we define $\alpha_2(t)=3\langle [\Delta\vec{r}'_1(t)]^4 \rangle /5 \langle [\Delta\vec{r}'_1(t)]^2 \rangle^2 - 1$.
In Fig. \ref{taua}(b), we plot $\tau_\text{ngp}$ as a function of $\dot{\gamma}$.
Similar to $\tau_\alpha$, $\tau_\text{ngp}$ also decreases with increasing $\dot{\gamma}$ as $\tau_\text{ngp} \sim \dot{\gamma}^{-0.65}$.
This behavior of $\tau_\text{ngp}$ was also experimentally observed in Ref. \cite{bessel_2007}.

\begin{figure}[t]
\begin{center}
\includegraphics[scale=1]{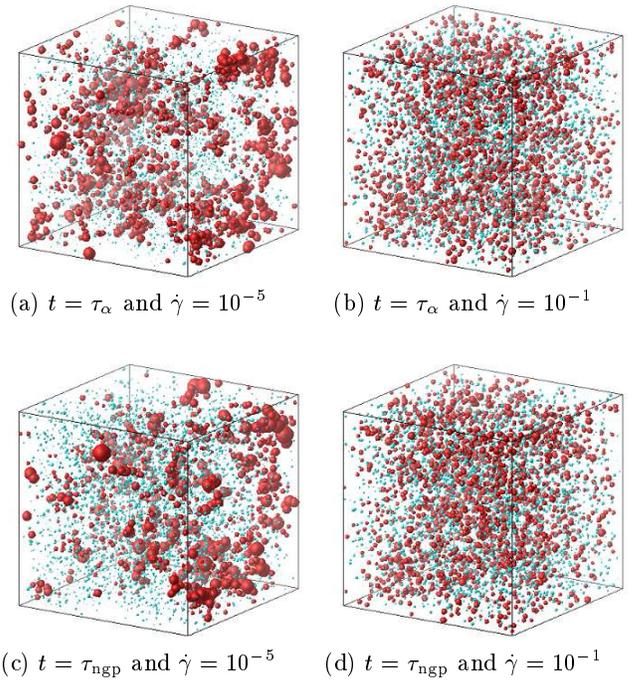}
\end{center}
\vspace*{0mm}
\caption{The visualization of the heterogeneous dynamics for Particle Species 1.
The temperature is $0.267$.
The time interval is $[t_0,t_0+\tau_\alpha] (t=\tau_\alpha)$ in (a) and (b) and $[t_0,t_0+\tau_\text{ngp}] (t=\tau_\text{ngp})$ in (c) and (d).
The shear rates are $10^{-5}$ in (a) and (c) and $10^{-1}$ in (b) and (d).
The radii of the spheres are $a_{1j}^2(t_0,t)$, and the centers are at $\vec{R}_{1j}(t_0,t)$. 
The red and blue spheres represent $a_{1j}^2(t_0,t) \ge 1$ (mobile particles) and
$a_{1j}^2(t_0,t)<1$ (immobile particles), respectively.}
\label{hetero}
\end{figure}

\subsection{Heterogeneous dynamics under shear flow}
First, we visualized the heterogeneous dynamics under a steady shear flow.
We show the spatial distribution of the particle mobility $a_{1j}^2(t_0,t)$ for Particle Species 1 in Fig. \ref{hetero}, where the particles are drawn as spheres with radii $a_j^2(t_0,t)$ located at $\vec{R}_j(t_0,t)$.
The time interval is $t=\tau_\alpha$ and $t=\tau_\text{ngp}$, and the temperature is $T=0.267$.
We demonstrate that the heterogeneity is significant at $\dot{\gamma}=10^{-5}$ but greatly weakened at $\dot{\gamma}=10^{-1}$ for both time intervals $t=\tau_\alpha$ and $\tau_\text{ngp}$.
Therefore, the strong shear flow suppresses the heterogeneous structure.
The same observation has also been observed in the heterogeneity of the bond breakage \cite{yamamoto1_1998} and of the ``overlapping'' function \cite{tsamados_2010,nordstrom_2011,heussinger_2010,heussinger2_2010,furukawa_2009}.

\begin{figure}
\begin{center}
\includegraphics[scale=1]{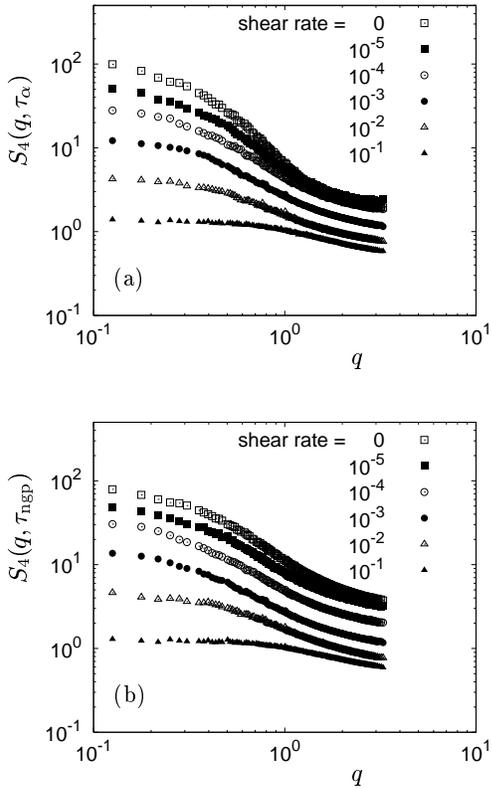}
\end{center}
\vspace*{-2mm}
\caption{
The spatial correlation function $S_{4}(q,t)$ for Particle Species 1.
The temperature is $T=0.267$.
The time interval $t$ is $\tau_\alpha$ in (a) and $\tau_{\text{ngp}}$ in (b).
The shear rate $\dot{\gamma}$ is  $0,\ 10^{-5},\ 10^{-4},\ 10^{-3},\ 10^{-2}$ and $10^{-1}$ from the highest curve to the lowest.
Note that $S_{4}(q,t)$ was calculated with a larger system $N = 10^5$.
}
\label{sdq}
\end{figure}

\begin{figure}
\begin{center}
\includegraphics[scale=1]{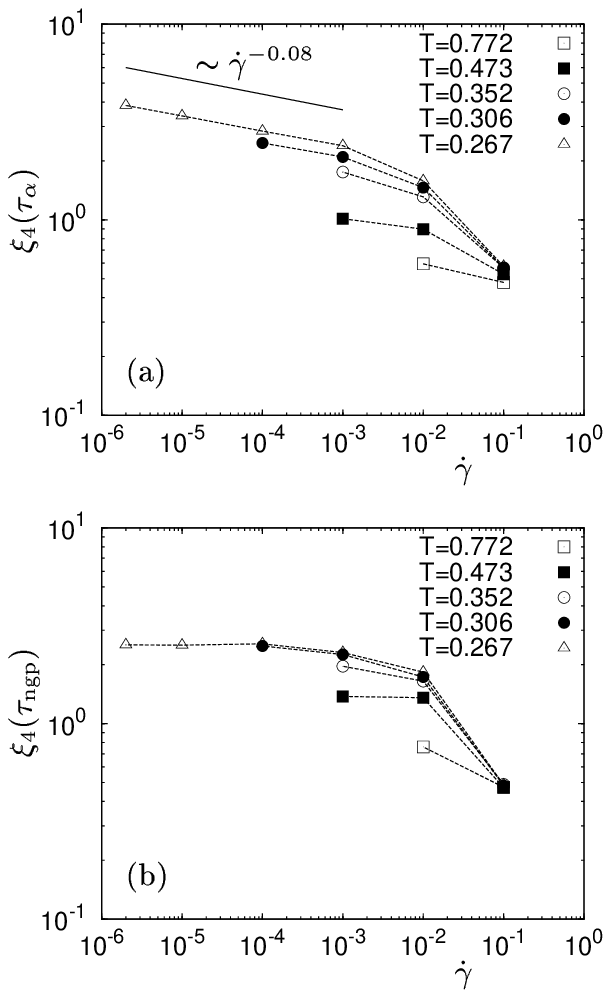}
\end{center}
\vspace*{-3mm}
\caption{The shear rate dependence of the correlation length $\xi_4(t)$ for various temperatures.
The time interval $t$ is (a) $\tau_\alpha$ and (b) $\tau_{\text{ngp}}$.}
\label{xi4}
\begin{center}
\includegraphics[scale=1]{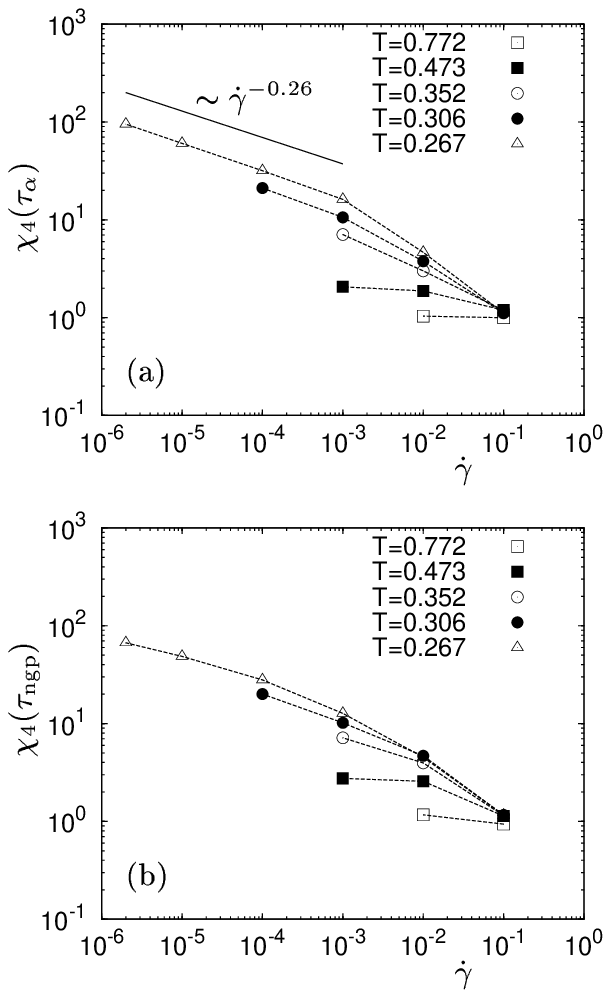}
\end{center}
\vspace*{-3mm}
\caption{The shear rate dependence of the intensity $\chi_4(t)$ for various temperatures.
The time interval $t$ is (a) $\tau_\alpha$ and (b) $\tau_{\text{ngp}}$.}
\label{chi4}
\end{figure}

\begin{figure}
\begin{center}
\includegraphics[scale=1]{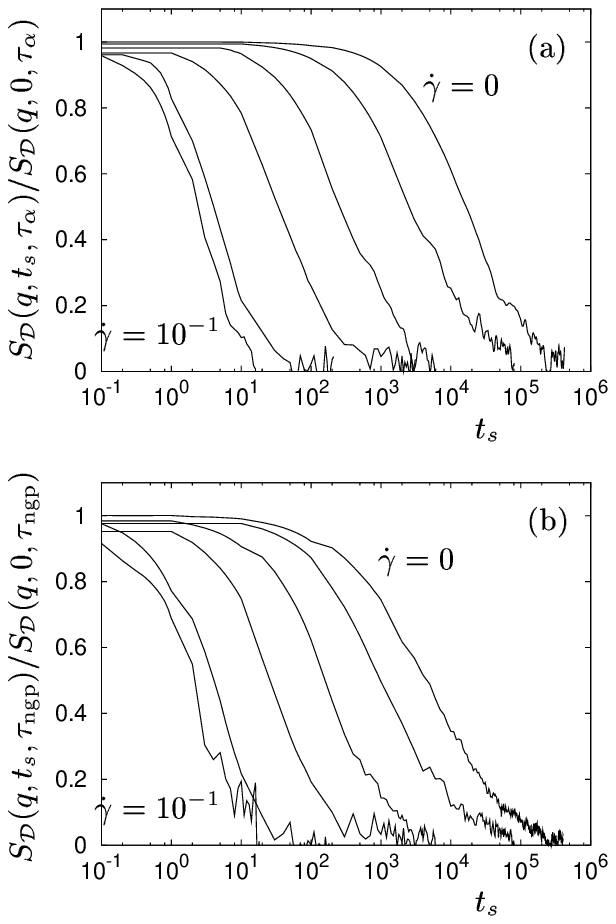}
\end{center}
\vspace*{-3mm}
\caption{The time decay of $S_{\vd}(q,t_s,t)$ for Particle Species 1 at $q=0.38$.
The time interval $t$ is $\tau_\alpha$ in (a) and $\tau_\text{ngp}$ in (b).
The temperature is $0.267$.
The shear rates $\dot{\gamma}$ are  $0,\ 10^{-5},\ 10^{-4},\ 10^{-3},\ 10^{-2}$, and $10^{-1}$ from right to left.}
\label{decay}
\begin{center}
\includegraphics[scale=1]{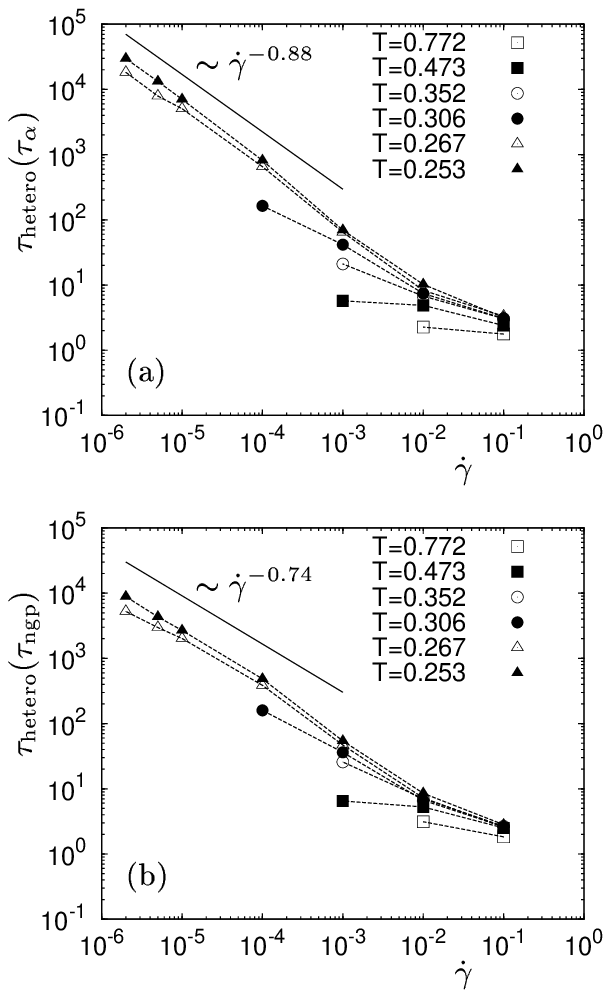}
\end{center}
\vspace*{-3mm}
\caption{The shear rate dependence of the lifetime $\tau_\text{hetero}(t)$ for various temperatures.
The time interval $t$ is (a) $\tau_\alpha$ and (b) $\tau_{\text{ngp}}$.}
\label{tauhetero}
\end{figure}

\subsection{Correlation length and intensity of dynamical heterogeneity}
Next, we quantified the correlation length $\xi_4(t)$ and the intensity $\chi_4(t)$ of the dynamical heterogeneity.
We calculated the spatial correlation function $S_4(q,t)$ defined in Eq. (\ref{sdqst}).
Note that we used a large system with $10^{5}$ particles for the calculation of $S_4(q,t)$.
Figure \ref{sdq} shows the wave number $q$ dependence of $S_{4}(q,t)$ for various shear rates $\dot{\gamma}$.
The temperature is $T=0.267$.
It can be observed that $S_{4}(q,t)$ at small wave numbers $q$ (long-distance scales) for $t=\tau_\alpha$ and $\tau_\text{ngp}$ decreases as the shear rate $\dot{\gamma}$ increases.
Therefore, the heterogeneous structure is weakened due to the steady shear flow as is also shown in Fig. \ref{hetero}.

We quantified the correlation length $\xi_4(t)$ and the intensity $\chi_4(t)$ by fitting the function $S_4(q,t)$ to the OZ form in Eq. (\ref{OZd}).
Note that $S_{4}(q,t)$ was carefully fitted to the OZ form in the range of $q \xi_{4}(t) < 1.5$ to obtain accurate values of $\xi_{4}(t)$ and $\chi_{4}(t)$ \cite{flenner_2010,flenner_2011}.
In Figs. \ref{xi4} and \ref{chi4}, we show the shear rate $\dot{\gamma}$ dependences of the correlation length $\xi_4(t)$ and the intensity $\chi_4(t)$ for various temperatures.
It is clear that both $\xi_4(t)$ and $\chi_4(t)$ decrease as $\dot{\gamma}$ increases for the time intervals $t=\tau_\alpha$ and $\tau_\text{ngp}$.
For the time interval $t=\tau_\alpha$, $\xi_4(\tau_\alpha)$ and $\chi_4(\tau_\alpha)$ can be scaled as $\xi_4(\tau_\alpha) \sim \dot{\gamma}^{-0.08}$ and $\chi_4(\tau_\alpha) \sim \dot{\gamma}^{-0.26}$ in Figs. \ref{xi4}(a) and \ref{chi4}(a), respectively.
The scaling component $2.6$ of $\chi_4(\tau_\alpha) \sim \dot{\gamma}^{-0.26}$ agrees well with $0.3$ in Refs. \cite{tsamados_2010,nordstrom_2011} but disagrees with $0.4-0.6$ in Ref. \cite{heussinger_2010}.

\subsection{Lifetime of dynamical heterogeneity}
Next, we quantified the lifetime $\tau_{\text{hetero}}(t)$ of the dynamical heterogeneity in terms of the correlation function ${S_{\vd}(q,t_s,t)}$ defined in Eq. (\ref{sdqt}).
We calculated ${S_{\vd}(q,t_s,t)}$ for the time intervals $t=\tau_\alpha$ and $\tau_{\text{ngp}}$.
Figure \ref{decay} shows the time decay of ${S_{\vd}(q,t_s,t)}$ at $q=0.38$ for various shear rates $\dot{\gamma}$.
Note that $q=0.38$ is the smallest waver number (the longest distance scale) that we used in calculating ${S_{\vd}(q,t_s,t)}$.
${S_{\vd}(q,t_s,t)}$ decays in the stretched exponential form with increasing the time separation $t_s$.
As shown in Fig. \ref{decay}, as the shear rate $\dot{\gamma}$ increases, ${S_{\vd}(q,t_s,t)}$ for both $t=\tau_\alpha$ and $\tau_\text{ngp}$ decays faster.

We quantified the lifetime $\tau_\text{hetero}(t)$ using the relaxation time of ${S_{\vd}(q,t_s,t)}$ as in Eq. (\ref{relaxsd}).
Figure \ref{tauhetero} shows $\tau_\text{hetero}(t)$ for $t=\tau_\alpha$ and $\tau_\text{ngp}$ as a function of $\dot{\gamma}$.
Similar to $\tau_\alpha$ and $\tau_\text{ngp}$, $\tau_\text{hetero}(t)$ also becomes small as $\dot{\gamma}$ becomes large.
Here, the scalings $\tau_\text{hetero}(\tau_\alpha) \sim \dot{\gamma}^{-0.88}$ and $\tau_\text{hetero}(\tau_\text{ngp}) \sim \dot{\gamma}^{-0.74}$ are observed.
The steady shear flow decreases the lifetime of the dynamical heterogeneity as well as the correlation length and the intensity.
By comparing three scalings $\xi_4 \sim \dot{\gamma}^{-0.08}$, $\chi_4 \sim \dot{\gamma}^{-0.26}$, and $\tau_\text{hetero} \sim \dot{\gamma}^{-0.88}$ for the time interval $t=\tau_\alpha$, we note that the lifetime is much more sensitive to the shear rate $\dot{\gamma}$ than the correlation length and the intensity.
Thus, the dynamical properties of the dynamical heterogeneity are altered much more by the shear flow than the static properties of the dynamical heterogeneity.

\begin{figure}
\begin{center}
\includegraphics[scale=1]{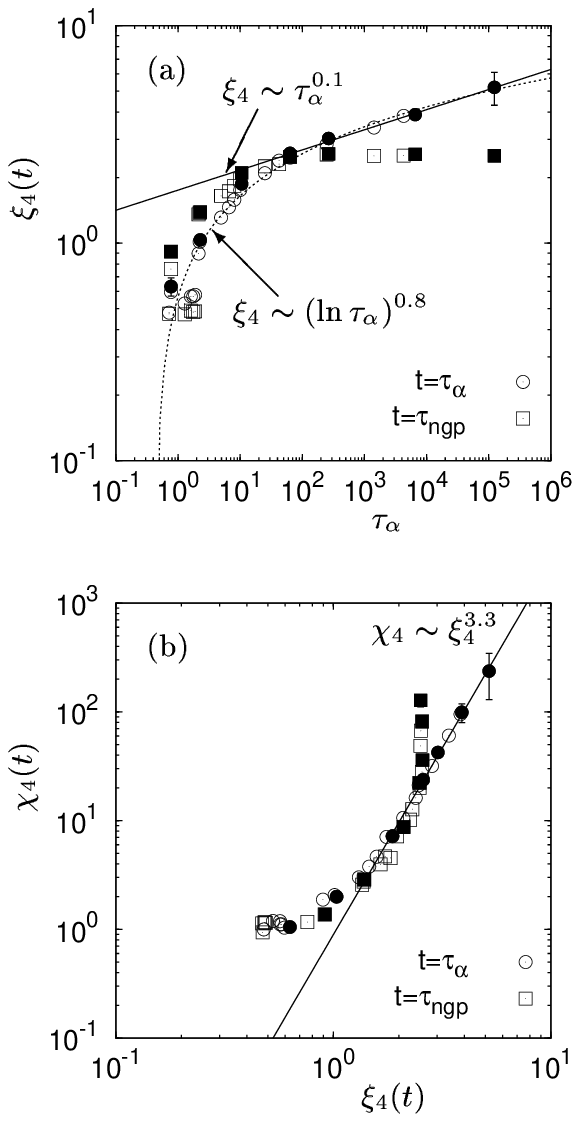}
\end{center}
\vspace*{-3mm}
\caption{
(a) The correlation length $\xi_4(t)$ versus the $\alpha$-relaxation time $\tau_\alpha$.
(b) The intensity $\chi_{4}(t)$ versus the correlation length $\xi_4(t)$.
The time intervals are $t=\tau_\alpha$ and $\tau_{\text{ngp}}$.
The white symbols are the values of $\xi_4(t)$ and $\chi_{4}(t)$ under shear flows.
The black symbols are the values in the equilibrium states from Ref. \cite{mizuno2_2010}.
The straight lines are power law fits: $\xi_4(\tau_\alpha) \sim \tau_\alpha^{0.1 \pm 0.01}$ in (a) and $\chi_4(\tau_\alpha) \sim \xi_4(\tau_\alpha)^{3.3 \pm 0.1}$ in (b).
The dashed curve is fit to $\xi_4(\tau_\alpha) \sim (\ln {\tau_\alpha})^{0.8 \pm 0.05}$.
}\label{tauaxi}
\begin{center}
\includegraphics[scale=1]{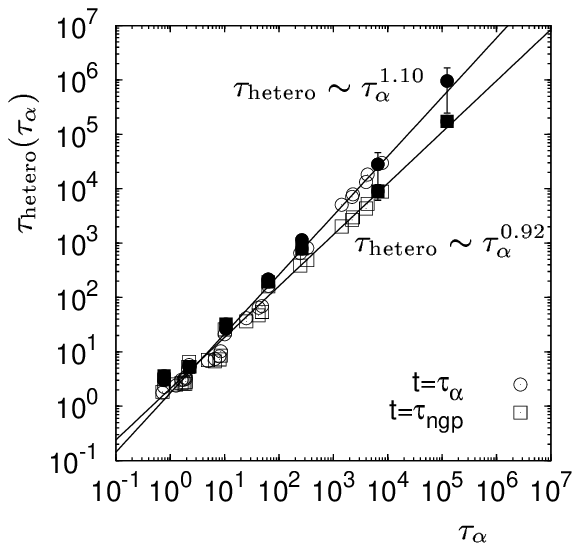}
\end{center}
\vspace*{-3mm}
\caption{The lifetime $\tau_{\text{hetero}}(t)$ versus the $\alpha$-relaxation time $\tau_\alpha$.
The time intervals are $t=\tau_\alpha$ and $\tau_{\text{ngp}}$.
The white symbols are the values of $\tau_{\text{hetero}}(t)$ under shear flows.
The black symbols are the values in the equilibrium states from Ref. \cite{mizuno_2010}.
The straight lines are power law fits: $\tau_{\text{hetero}}(\tau_\alpha) \sim \tau_{\alpha}^{1.10 \pm 0.02}$ and $\tau_{\text{hetero}}(\tau_{\text{ngp}}) \sim \tau_{\alpha}^{0.92 \pm 0.02}$.
}\label{tauatauh}
\end{figure}

\subsection{Mapping concept for correlation length, intensity, and lifetime of dynamical heterogeneity}
Finally, we demonstrated that three quantities, the correlation length $\xi_4(t)$, the intensity $\chi_4(t)$, and the lifetime $\tau_\text{hetero}(t)$, can be mapped onto those in the equilibrium state.
In Fig. \ref{tauaxi}(a), we plot $\xi_4(t)$ versus $\tau_\alpha$ for both time intervals $t=\tau_\alpha$ and $\tau_\text{ngp}$.
In the same figure, we also plot the values in the equilibrium state.
It can be observed that the values $\xi_4(t)$ for $t=\tau_\alpha$ and $\tau_\text{ngp}$ collapse onto a single curve as a function of $\tau_\alpha$.
This result means that the correlation length $\xi_4(t)$ in the sheared state can be mapped onto that in the equilibrium state through the $\alpha$-relaxation time $\tau_\alpha$.
For the time interval $t=\tau_\alpha$, we obtained the scaling relationships between $\xi_{4}(\tau_\alpha)$ and $\tau_\alpha$: $\xi_{4}(\tau_\alpha) \sim \tau_\alpha^{0.1}$ (a power law fit) or $\xi_4(\tau_\alpha) \sim (\ln {\tau_\alpha})^{0.80}$ (a slower-than-power law fit).
These fits are exactly the same as the equilibrium fits within the numerical errors \cite{mizuno2_2010}.
In addition, in Fig. \ref{tauaxi}(b), we show $\chi_4(t)$ versus $\xi_4(t)$ for $t=\tau_\alpha$ and $\tau_\text{ngp}$.
From Fig. \ref{tauaxi}(b), similarly to $\xi_4(t)$, $\chi_4(t)$ also collapses onto a single curve as function of $\xi_4(t)$.
Therefore, the values $\chi_4(t)$ in the sheared state can also be mapped onto those in the equilibrium state.
For the time interval $t=\tau_\alpha$, we obtained the scaling relationship between $\chi_{4}(\tau_\alpha)$ and $\xi_4(\tau_\alpha)$. It is a power law fit $\chi_{4}(\tau_\alpha) \sim \xi_4(\tau_\alpha)^{3.3}$, which is same as the equilibrium fit within the numerical errors \cite{mizuno2_2010}.
It should be noted that $\tau_\alpha$ decreases with increasing $\dot{\gamma}$ as $\tau_\alpha \sim \dot{\gamma}^{-0.8}$, which leads to $\xi_4(\tau_\alpha) \sim \tau_\alpha^{0.1} \sim \dot{\gamma}^{-0.08}$ and $\chi_4(\tau_\alpha) \sim \xi_4(\tau_\alpha)^{3.3} \sim \dot{\gamma}^{-0.26}$.
These scaling relations coincide with the observations in Figs. \ref{xi4}(a) and \ref{chi4}(a), respectively.

Furthermore, figure \ref{tauatauh} shows the lifetime $\tau_{\text{hetero}}(t)$ versus $\tau_\alpha$ for $t=\tau_\alpha$ and $\tau_\text{ngp}$.
This figure shows that $\tau_\text{hetero}(\tau_\alpha)$ and $\tau_\text{hetero}(\tau_{\text{ngp}})$ fall into the lines $\tau_{\text{hetero}}(\tau_\alpha) \sim \tau_{\alpha}^{1.10}$ and $\tau_{\text{hetero}}(\tau_{\text{ngp}}) \sim \tau_{\alpha}^{0.92}$, similarly to $\xi_4(t)$ and $\chi_4(t)$.
These lines are exactly same as the equilibrium ones within the numerical errors \cite{mizuno_2010}.
We note that $\tau_{\text{hetero}}(\tau_\alpha) \sim \tau_{\alpha}^{1.10} \sim \dot{\gamma}^{-0.88}$ and $\tau_{\text{hetero}}(\tau_{\text{ngp}}) \sim \tau_{\alpha}^{0.92} \sim \dot{\gamma}^{-0.74}$ coincide with the observation in Fig. \ref{tauhetero}.
Therefore, not only the correlation length and the intensity but also the lifetime can be mapped onto those in the equilibrium state through the $\alpha$-relaxation time.
As we mentioned, previous studies \cite{yamamoto_1998,yamamoto1_1998,haxton_2007} have found that the material properties, such as the shear viscosity $\eta$ and the diffusion constant $D$, under steady shear flow can be a single function of the $\alpha$-relaxation time $\tau_\alpha$ or the effective temperature $T_{\text{eff}}$.
Our results demonstrated that the mapping concept is valid for the material properties (the shear viscosity and the diffusion constant) as well as the properties of dynamical heterogeneity (the correlation length, the intensity, and the lifetime).

\section{Conclusion}  \label{3conclusion}
In conclusion, we examined the dynamical heterogeneity in a supercooled liquid under sheared conditions.
Dynamical heterogeneity can be characterized by three quantities: the correlation length $\xi_4(t)$, the intensity $\chi_4(t)$, and the lifetime $\tau_\text{hetero}(t)$ .
Using the correlation functions of the particle dynamics, we quantified all three quantities for two time intervals $t=\tau_\alpha$ and $\tau_\text{ngp}$.
Our results demonstrated that not only the correlation length and the intensity but also the lifetime decrease as the shear rate $\dot{\gamma}$ increases.
Therefore, the steady shear flow suppresses the length scale as well as the time scale of the dynamical heterogeneity.
For the time interval $t=\tau_\alpha$, three quantities can be scaled as $\xi_4 \sim \dot{\gamma}^{-0.08}$, $\chi_4 \sim \dot{\gamma}^{-0.26}$, and $\tau_\text{hetero} \sim \dot{\gamma}^{-0.88}$, which indicates that, due to the steady shear flow, the lifetime decreases much more than the correlation length and the intensity.

Furthermore, we numerically demonstrated that the values of $\xi_4(t)$, $\chi_4(t)$, and $\tau_\text{hetero}(t)$ collapse onto a single curve as function of the $\alpha$-relaxation $\tau_\alpha$ similarly to the material properties such as the viscosity $\eta$ and the diffusion constant $D$.
For the time interval $t=\tau_\alpha$, we obtained exactly the same scaling relations between the three quantities as the equilibrium ones, i.e., $\xi_{4} \sim \tau_\alpha^{0.1}$ (or $\xi_4 \sim (\ln {\tau_\alpha})^{0.80}$), $\chi_{4} \sim \xi_4^{3.3}$, and $\tau_{\text{hetero}} \sim \tau_{\alpha}^{1.10}$.
Our results indicate that three quantities of the dynamical heterogeneity can be mapped onto those in the equilibrium state, i.e., the mapping concept for three quantities is valid.

In the present study, we did not acknowledge the anisotropy of the dynamical heterogeneity, and the values of $\xi_4(t)$, $\chi_4(t)$, and $\tau_\text{hetero}(t)$ were averaged over the direction of the wave vector $\vec{q}$.
However, as we mentioned, the dynamical heterogeneity becomes anisotropic in the sheared situation, and the anisotropic dynamics can be related to shear-thinning \cite{furukawa_2009}.
How the dynamical heterogeneity and its three quantities depend on the direction of the wave vector $\vec{q}$ is interesting and important subject.

%

\end{document}